*International*

*Virtual*

*Observatory*

*Alliance*

# IVOA Photometry Data Model

# Version 1.0

## *IVOA Recommendation*

## *2013 Oct 05*

**Editor(s):**
Jesús Salgado
Pedro Osuna

**Author(s):**
Jesús Salgado
Carlos Rodrigo
Pedro Osuna
Mark Allen
Mireille Louys
Jonathan McDowell
Deborah Baines
Jesús Maíz Apellániz
Evanthia Hatziminaoglou
Sebastien Derriere
Gerard Lemson

# Abstract

The Photometry Data Model (**PhotDM**) standard describes photometry filters, photometric systems, magnitude systems, zero points and its interrelation with the other IVOA data models through a simple data model. Particular attention is given necessarily to optical photometry where specifications of magnitude systems and photometric zero points are required to convert photometric measurements into physical flux density units.

.

# Status of This Document

*The first release of this document was 2010 May 05.*

*This document has been produced by the IVOA DM Working Group.*

*It has been reviewed by IVOA Members and other interested parties, and has been endorsed by the IVOA Executive Committee as an IVOA Recommendation. It is a stable document and may be used as reference material or cited as a normative reference from another document. IVOA's role in making the Recommendation is to draw attention to the specification and to promote its widespread deployment. This enhances the functionality and interoperability inside the Astronomical Community.*

*A list of current IVOA Recommendations and other technical documents can be found at http://www.ivoa.net/Documents/.*

# Acknowledgements

We acknowledge the EuroVO Science Advisory Committee for the review of the initial versions of the document and to the developers who have contributed to the data model reference implementations.

# Contents

# 1 Introduction

A key role of the VO is to help astronomers find data and to combine that data in a scientifically meaningful way. A Spectral Energy Distribution (SED) is an example of combining data whereby flux density measurements of an astrophysical source at different spectral energy coordinates (wavelengths/frequencies/energy) [1][2][3][4] are plotted as a graph of flux density versus a spectral energy coordinate. SEDs that cover a wide range of the electromagnetic spectrum are particularly useful for identifying the underlying physical processes operating in the astrophysical source, and the use of SEDs is becoming more prevalent as astronomy takes an increasingly multi-wavelength approach. To combine individual flux density measurements and their spectral energy coordinates into an SED, these 'photometry' measurements must be described in sufficient detail to allow for the conversion to compatible flux density and spectral energy units, taking into account the nature of the spectral energy bandpass of the measurements, as well as the apertures and other details of the measurements. This document outlines a photometry data model to describe photometric measurements in a standard way.

The photometry data model aims to describe the essential elements of flux density measurements made within all spectral energy domains across the electromagnetic spectrum. In some domains this is relatively straight forward, such as in radio astronomy where measurements are commonly expressed in flux density units, and where data are readily combined into SEDs. The data model fields required to describe such a radio flux density measurement includes a specification of the bandpass, the units of the measurement and the associated uncertainties. Optical photometry measurements are however commonly expressed in magnitudes, and a greater level of description of the magnitude systems and bandpasses are required to support the conversion of these measurements into flux densities that could be combined into an SED. As such, much of this document is necessarily devoted to defining the data model fields required to describe optical photometry measurements.

Astronomical flux density measurements will often require a greater level of description than provided by this simple model. The level of accuracy required depends strongly on the scientific use of the data. A study of broadband SEDs of active galaxies may tolerate 20% uncertainties in the flux density measurements, and it is usually sufficient in these cases to use average values for the spectral energy coordinates of the bandpasses. Fitting to stellar models or science that employs photometric measurements to derive photometric redshifts requires a much greater level of accuracy. To manage the different levels of description we take the overall approach that the photometry data model should include the most generic elements required to describe photometric measurements, and that the photometry data model is intended to be used in coordination with the IVOA Spectrum Data model and the IVOA Characterization Data Model.

The scientific use case that has guided the choice of the level of description of the metadata fields in the Photometry Data Model is the use of the large collections of photometric data that are published in catalogues (e.g. Vizier, http://vizier.u-strasbg.fr/) in SEDs. The Photometry Data Model provides the metadata fields for describing the photometry measurements in catalogues, so that those data could then be added to, or compared with an SED.

The intended practical use of the Photometry Data Model is that the metadata fields defined here will be included in the metadata of catalogues, or of photometry data stored as a pseudo-spectrum. These data would then be made accessible using Simple Spectral Access Protocol (SSAP) or Table Access Protocol (TAP) services so that the photometric measurements can be used and combined in scientific software tools.

The proposed model is based on the description of the photometry filters, and the description of how the units of the measurement are related to flux density. The photometry filter description may be as simple as specifying a central spectral energy coordinate and a bandpass width. The more detailed description of optical bandpasses is supported by allowing for specification of filter transmission curves, and the photometric zero points necessary for the conversion of magnitudes to flux densities.

Information on the properties of filters is not always easily available, and is sometimes only specified in manuals or in the literature and often not in digital form. To aid the use of filters information, in particular as part of the Photometry Data Model metadata fields, we propose a mechanism for referencing external filter information. Such a *Filter Profile Service* would expose this information so software client applications could discover it. This document proposes a standardization of a protocol to be used by Filter Profile Services.

The following sections of this document summarize some key points about astronomical photometry (Section 2). The detailed metadata structure of the data model is presented in Section 3. Section 4 describes use cases in which the model description could be used in making photometry data available through VO protocols, and, very briefly, how scientific tools could use this information.

The figure below shows where Photometry DM fits within the IVOA architecture:

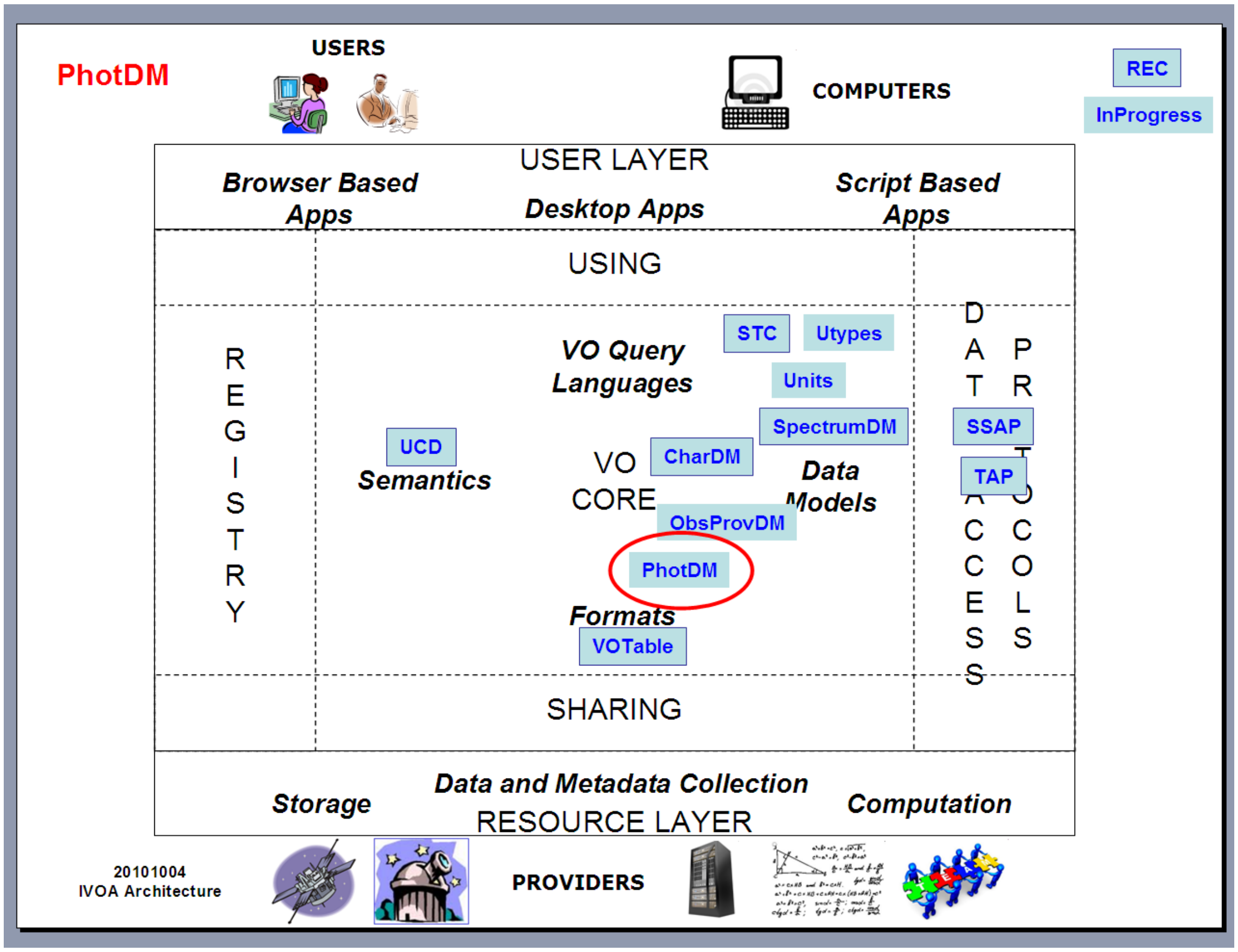

PhotDM is related to other IVOA Data models (Spectrum DM, Characterization DM, Observation and Provenance DM), and is intended to provide photometry metadata for data that would be accessed via the IVOA Data Access Protocols such as SSAP (Simple Spectra Access Protocol) or TAP (Table Access Protocol).

As with most of the VO Data Models, PhotDM makes use of STC, Utypes, Units and UCDs. PhotDM will be serializable with a VOTable

## 2 Astronomical Photometry

Astronomical photometry refers to measuring the brightness, flux or intensity of an astrophysical object. Consider an astronomical source with a flux density at the observer *F(x)*, where *x* is a spectral coordinate (wavelength, frequency or energy). The photometry measurement will be related to <*F*> a flux weighted integral of this flux density over an observed band with a relative spectral response *T(x).* The flux weighted integral in its most simple form is defined as

$$<F> = \int F(x)T(x)dx$$

Calibration of photometric measurements is in general done by comparison to a reference spectrum that has a known effective flux density $f_0$ at a specific spectral band.

For this reference spectrum, the flux weighted integral is defined as:

$$<F_R> = \int F_R(x)T(x)dx$$

so that the effective flux density of the source can be evaluated as:

$$f = f_0 \left( \frac{<F>}{<F_R>} \right)$$

This represents the most simple and easy to use flux measurement. Flux measurements expressed in physical flux density units can be easily combined into SEDs. Many flux measurements published in catalogues of radio sources for example are simple flux densities of this form.

In optical photometry measurements are often expressed as magnitudes and it is necessary to define the magnitude system being used, and the zero point fluxes of the reference spectrum.

Pogson magnitudes are defined as:

$$m = -2.5\log_{10}(<F>)$$

which when compared to a reference spectrum $F_R$ leads to:

$$m = m_R - 2.5\log_{10}(<F>/<F_R>)$$

As explained above, this is equivalent to:

$$m = m_R - 2.5\log_{10}(f / f_0)$$

so that

$$f = f_0 10^{-0.4(m-m_R)}$$

Using this expression a measurement in magnitudes can be converted to a flux density, given the zero point flux of the reference spectrum. The magnitude of reference $m_R$ and the zero point $f_0$ will be defined in the document. $m_R$ is most often chosen to be zero (or one for linear photometric systems) in most of the photometric systems although, usually, continuous recalibration of the photometric system usually produces a deviation of this value.

There are a number of magnitude systems that are defined by the reference spectrum. The three most commonly used magnitude systems are the Vega magnitude, $AB_\nu$ magnitude and $ST_\lambda$ magnitude systems. The Vega magnitude system uses the spectrum of Vega (Alpha Lyrae) as the reference spectrum $F_R(x)$. The $AB_\nu$ magnitude system uses reference spectrum defined by a constant flux density per unit frequency ($F_\nu$ ) and the $ST_\lambda$ magnitude system uses a reference spectrum of a constant flux density per unit wavelength ($F_\lambda$). The values of $F_\nu$ and $F_\lambda$ that respectively define the zero points $m_{AB,\nu}=0$ and $m_{ST,\lambda}=0$ have been chosen to be the mean flux density of Vega in the Johnson V band.

$$m_{AB,\nu} = 0$$
$$f_\nu = 3.63 \times 10^{-20}\, erg\, cm^{-2}\ s^{-1}\ Hz^{-1}$$

$$m_{ST,\lambda} = 0$$
$$f_\lambda = 3.63 \times 10^{-9}\, erg\, cm^{-2}\ s^{-1}\text{Å}^{-1}$$

A convenient graphical representation of these systems is shown in Figure 3.1 of the Synphot user's manual: (http://www.stsci.edu/resources/software_hardware/stsdas/synphot/SynphotManual.pdf).

For a photometric system that uses Vega magnitudes, the zero point flux for each filter is the average flux density of Vega over that bandpass $f_{Vega}$. Some typical values of $f_{Vega}$ are tabulated in [15] for the Johnson photometric system. Although the agreed Vega spectrum has changed historically, the commonly referred to spectrum of Vega in digital form described in (Bohlin, Gilliland) [16] is available as file at alpha_lyr_stis_002.fits at:

http://www.stsci.edu/instruments/observatory/cdbs/calspec

In the AB system, the flux density (in units of *erg* $cm^{-2}$ $s^{-1}$ $Hz^{-1}$) corresponding to a given magnitude is simply obtained via:

$$f_\nu = 10^{-0.4(m_{AB}+48.6)}$$

And, in the same way, in the ST system, the flux density (in units of *erg* $cm^{-2}$ $s^{-1}$ $Å^{-1}$) corresponding to a given magnitude is:

$$f_\lambda = 10^{-0.4(m_{ST}+21.1)}$$

Another magnitude system is the Asinh magnitude system in which magnitudes are defined as

$$m = \frac{2.5}{\ln(10)}\left[\sinh^{-1}\left(\frac{f}{2bf_0}\right) + \ln(b)\right]$$

where b is known as the softening parameter. Details of the Asinh magnitude system and the softening parameters are described in http://www.sdss.org/DR7/algorithms/photometry.html

# 3 Photometry Data Model

The model shown in Figure 1, organizes the structure and detailed metadata fields of the Photometry Data Model in a logical manner, and shows the relationship to other IVOA data models. The metadata fields for each class specify the essential elements required to describe a photometric measurement.

The main class in this diagram is Photometry Filter. This class contains all the attributes necessary to describe a filter from the data discovery point of view.

A Photometric System is a grouping of individual Photometry Filters. This may represent a particular set of filters that are related in some way.

A magnitude system is characterized for a certain reference spectrum that will produce a certain zero point for a certain photometry filter. This reference spectrum could be an ideal one (as in STmag and ABmag systems), a Vega-like spectrum (as in Vegamag systems) (please notice that different Vega spectrum versions have been historically used) or any other. In many cases, the reference spectrum has been calculated as an average of spectra from several astronomical objects. This would be characterized by a set of Source instances.

A zero point would then be a flux value that can be considered as zero magnitude, so its value will allow conversions from fluxes to magnitudes and the other way around. It has associated a photometry filter and it also depends on the magnitude system (reference spectrum) used to calculate this magnitude.

There are different types of zero points (Pogson, asinh, linear etc) that will essentially differ in the way that getFluxFromMagnitude and getMagnitudeFromFlux operators are implemented plus extra information that could be needed to do these conversions.

An intermediate class, PhotCal, can be understood as a certain photometry filter instance, i.e., a certain photometry filter using a certain magnitude system and linked to a certain zero point class. PhotCal is the class node where SpectralDM v2.0 interacts. It binds the filter and zeropoint information to the Flux Axis calibration in SpectralDM v2.0. It can be understood as the calibration configuration used , bringing together a specific photometry filter instance with magnitude system and zeropoint. It leverages the handling of photometric data through IVOA protocols e.g. SSAP or TAP services.

A Spectrum would have a Characterization Coordsys element that will have associated a certain PhotCal element in the case of photometry data. Using this information, magnitudes from different photometric systems could be compared between them or compared to spectroscopic data expressed in flux.

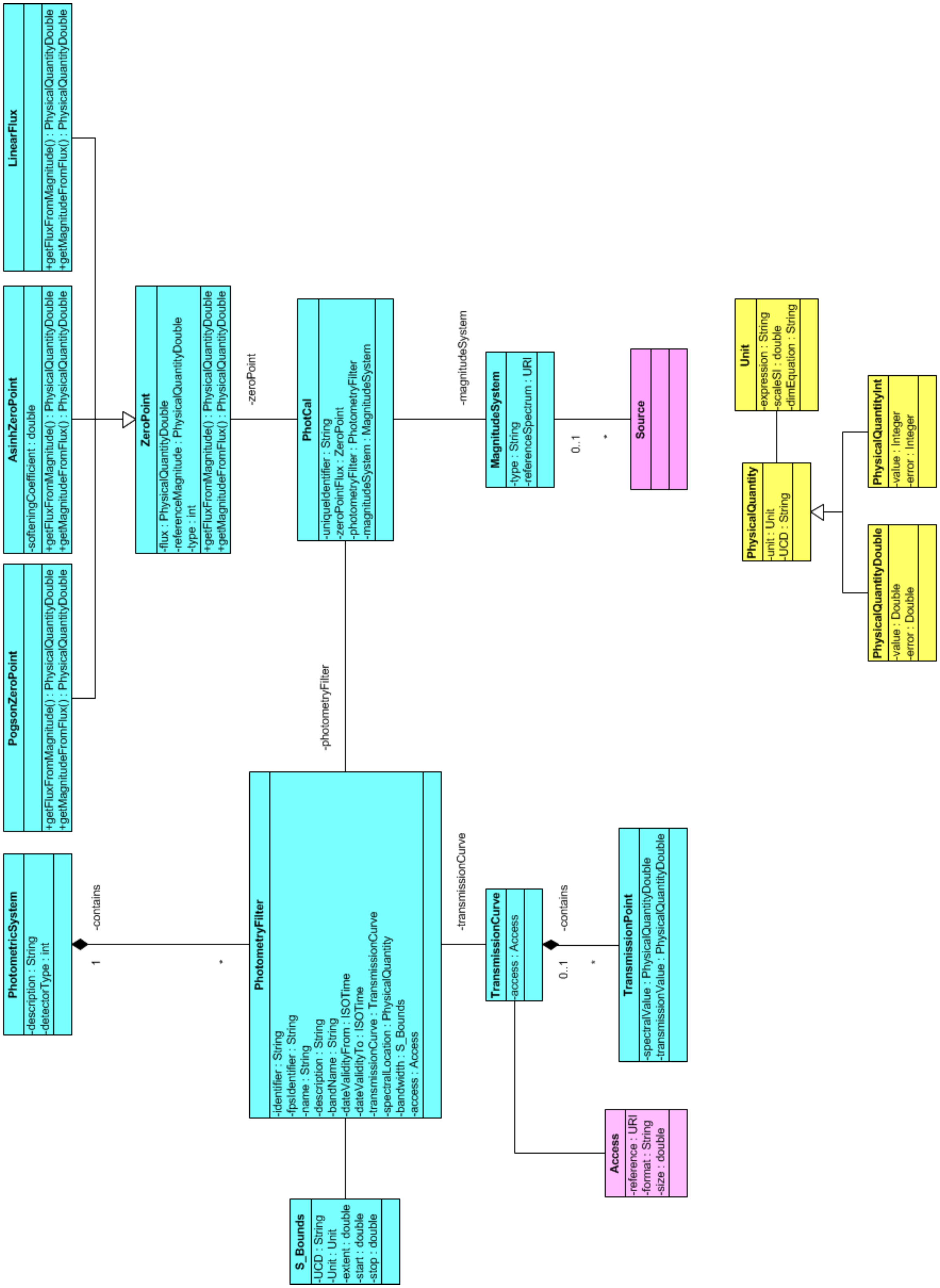

**Figure 1 In blue, class diagram of the Photometry Calibration Data Model: reused classes from other IVOA DMs are shown in pink (SourceDM not standardized at IVOA level yet). In yellow, simplified physical quantity class that will be included in the IVOA profile to glue the different fields that describe a measurement.**

In order to fully describe values of the magnitudes inside photometry point instances, the class diagram makes use of physical quantity classes. These classes glue all the basic fields that compose a physical measurement: value, error, units, etc. However, within the present specification, we will describe individual attributes of the different quantities and as a consequence. All the utypes will be also generated from individual physical quantity attributes what will facilitate the use within IVOA Data Access Layer protocols.

## *3.1 PhotometricSystem Class*

This class briefly describes the photometric system that contains a set of photometry filters. Photometry filters can be contained in a certain photometric system as part of the same observatory/telescope or as part of a known system.

### 3.1.1 PhotometricSystem.description: String

This String contains a human readable short-text representation of the photometric system. This will allow client applications to display textual information to final users.

Examples:

```
Sloan
Johnson
```

### 3.1.2 PhotometricSystem.detectorType: integer

Detector type associated to this photometric system. Possible values are:

| Type of detector | Value | Examples |
|---|---|---|
| Energy Counter | 0 (default) | Energy amplifiers devices |
| Photon Counter | 1 | CCDs or photomultipliers |

This will be used in order to decide how to calculate the flux average in, e.g., the synthetic photometry calculations. At current state, this list is exhaustive. See photometry filter transmission curve description to understand how to use this field.

## *3.2 PhotometryFilter Class*

This is the main class that describes a photometry filter.

### 3.2.1 PhotometryFilter.identifier: String

This field identifies, in a unique way, within a certain Photometry Filter Profile service, a filter. Although the main requirement of this data model field is to be unique within a Filter Profile Service, the suggested syntax would be:

```
Facility/Subcategory/Band[/Suffix]
```

where *Facility* is the telescope, observatory, space mission, etc that has this filter, *Subcategory* is a meaningful classification of filters within a facility (usually instrument), *Band* is the generic name used to describe the wavelength band used by this filter and *Suffix* is optional metadata added to the unique identifier string to ensure uniqueness within a Filter Profile Service.

Example:

```
SDSS/SDSS.G/G
```

### 3.2.2 PhotometryFilter.fpsIdentifier: String

IVOA identifier of the filter profile service where this photometry filter is registered to be used in the discovery of all the relevant photometry filter properties.

This identifier follows the IVOA syntax defined for IVOA identifiers [5] which gives a string built up as:

```
ivo ://<ivoa authority id>/<resource key>
```

Example:

```
ivo://svo/fps
```

where svo is the authority id, fps is the resource key of the service.

The service url of the filter profile service would be obtained from the registry by requesting the associated information of this registry resource, e.g., once registered the service URL associated to this Filter Profile Service would be, e.g.:

```
http://svo.cab.inta-csic.es/theory/fps/
```

### 3.2.3 PhotometryFilter.name: String

This String contains a human readable representation of the filter name. This will allow client applications to display information to the final user.

Example:

```
SDSS.G
```

### 3.2.4 PhotometryFilter.description: String

This String contains a verbose human readable string description of the filter. This will allow client applications to display text information to the final user.

### 3.2.5 PhotometryFilter.bandName: String

This String contains a standard representation of the spectral band associated to this filter (if any). This information is useful for human interpretation but it is discourage to use it for discovery purposes. The reason is that a filter is not always properly represented by a standard band so filters could be lost in a query response.

Examples:

```
U
B
V
```

Where U,B,V corresponds to ultraviolet, blue and visible respectively.

### 3.2.6 PhotometryFilter Time Validity Range

The following fields will be used to characterize the validity range of this specific photometry filter configuration. This is particularly useful for ground based telescopes where filter, electronics, etc could easily change generating versions of the same photometry filter.

#### 3.2.6.1 PhotometryFilter.dateValidityFrom: ISOTime

Start time of the time coverage when this filter configuration is applicable. String time format accepted, ISO8601:

```
YYYY-MM-DD[T[hh[:mm[:ss[.s]]]]]
```

#### 3.2.6.2 PhotometryFilter.dateValidityTo: ISOTime

End time of the time coverage when this filter configuration is applicable. String time format accepted, ISO8601:

```
YYYY-MM-DD[T[hh[:mm[:ss[.s]]]]]
```

### 3.2.7 PhotometryFilter.transmissionCurve

Here we consider how wavelengths/frequencies are filtered in the whole acquisition chain for a calibrated observation stemming from a given data collection.

This means that within the same data collection most observations will point to the same PhotometryFilter.transmissionCurve.

The effective transmission curve may be represented as a 2-D graph that describes the transmission properties of the filter over a wavelength range defined by the filter bandpass.

It is composed of a spectral coordinate in the x-axis and a scalar in the y-axis. This effective response curve encloses all the possible components that modifies the energy/photon collection, including detector, telescope and even atmosphere for transmission curves referenced in measurements. Most modern surveys try to reduce everything according to a given airmass (e.g. 1.3) and this is particularly important for ground-based filters with $\lambda < 4000\ \text{Å}$ or $\lambda > 7000\ \text{Å}$

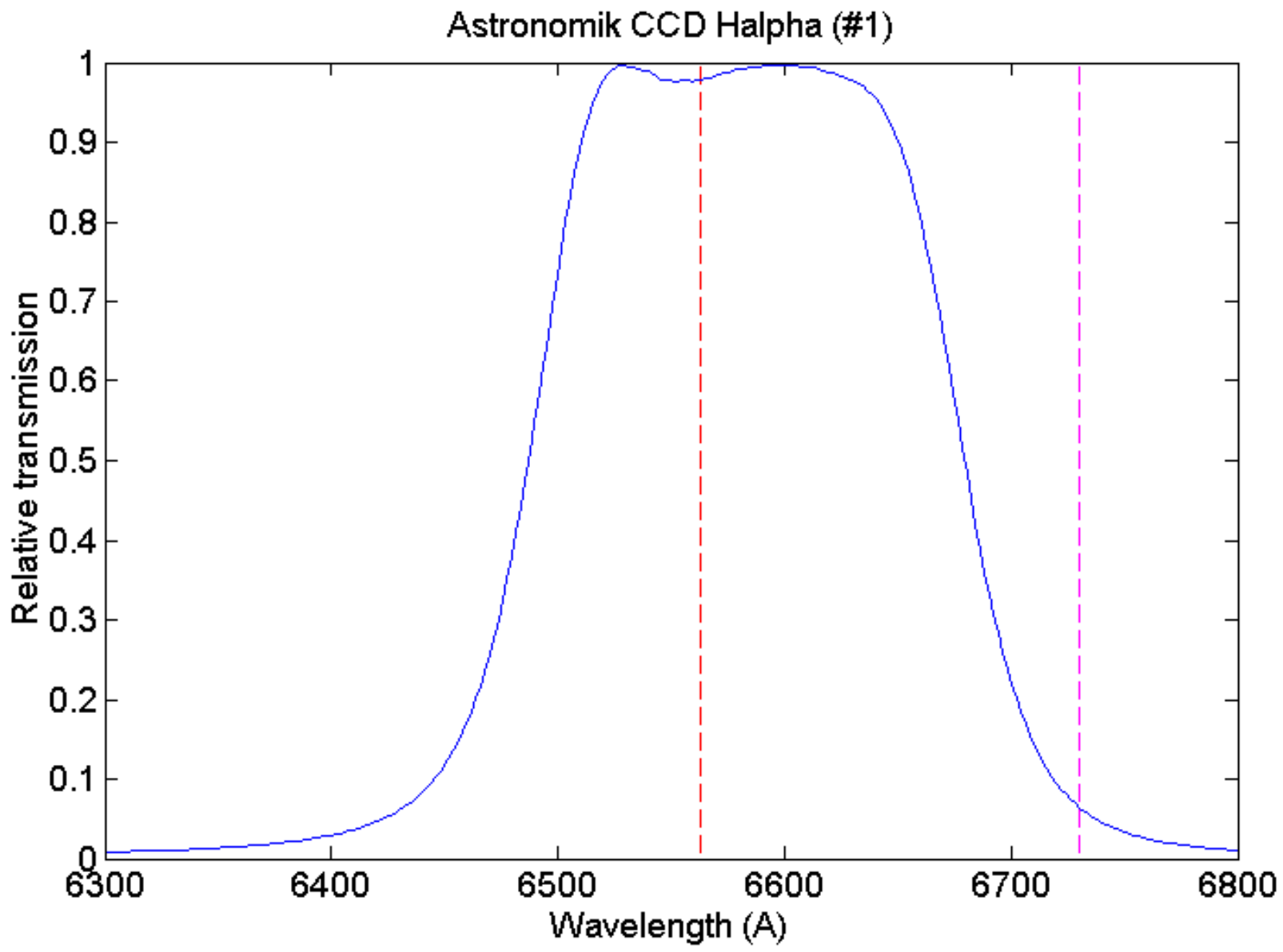

**Figure 2: Transmission Curve example**

This curve can be used, e.g. for the creation of synthetic photometry [8][9] from an observational or a theoretical spectrum by applying it to the spectrum in the filter band-pass. Taking as input a certain flux, the effective flux as seen using a certain filter would be, for energy counters [10]:

$$f(\lambda_{eff}) = \frac{\int T(\lambda)F(\lambda)d\lambda}{\int T(\lambda)d\lambda}$$

And for photon counters (like CCDs or photomultipliers):

$$f(\lambda_{eff}) = \frac{\int T(\lambda)F(\lambda)\lambda d\lambda}{\int T(\lambda)\lambda d\lambda}$$

Where $T(\lambda)$ is the transmission curve, $f(\lambda)$ is the flux of the spectrum. As the transmission curve is defined only in the filter band-pass, the limits of the integrals corresponds to the spectral range where the transmission curve is defined (stored as *PhotometryFilter.bandwidth* in this data model)

The transmission curve can be closely (although not fully) identified as an array of points as in a spectrum. There are various ways to provide this information either directly in an embedded table, or using a reference to a serialized table file.

Spectral and transmission coordinates can be gathered directly as a table using TransmissionPoint utypes (see 3.3.3)..

### 3.2.8 PhotometryFilter.spectralLocation.value: double

A spectral coordinate value that can be considered by the data provider as the most representative for this specific filter band-pass. The selection of this value should take into account the filter transmission curve profile and in general should be close to the wavelength mean value, defined [7] as:

$$\lambda_{mean} = \frac{\int T(\lambda)\lambda d\lambda}{\int T(\lambda)d\lambda}$$

where $\lambda_{mean}$ is the spectral bounds mean value, $T(\lambda)$ is the transmission curve (see below), $\lambda$ is the wavelength. Please notice that, since the transmission curve will only be defined in a specific spectral range, the integrals will also be effectively defined in this spectral range.

Another convenient definition of an effective wavelength is the “pivot wavelength” defined as follows:

$$\lambda_{pivot} = \sqrt{\frac{\int T(\lambda)\lambda d\lambda}{\int T(\lambda)\, d\lambda / \lambda}}$$

It can be proved that the pivot wavelength fulfills the following relation between the $f_\lambda$ and $f_\nu$:

$$\langle f_\nu \rangle = \langle f_\lambda \rangle \lambda^2_{pivot} / c$$

Other definitions for effective wavelengths commonly used in the literature are source dependent as, e.g., the isophotal wavelength:

$$\lambda_{mean} = \frac{\int \lambda F_\lambda(\lambda) T(\lambda) d\lambda}{\int F_\lambda(\lambda) T(\lambda) d\lambda}$$

Or the photon distribution based effective wavelength:

$$\lambda'_{mean} = \frac{\int \lambda^2 F_\lambda(\lambda) T(\lambda) d\lambda}{\int \lambda F_\lambda(\lambda) T(\lambda) d\lambda}$$

but these source dependent definitions have two caveats:

- Real spectra do not necessarily satisfy the requirements of the mean value theorem, which could produce multiple values for the wavelength
- Calculation of these wavelengths implies the knowledge of $F_\lambda$ (usually what you want to measure) and it does not look like an intrinsic property of the photometry filter.

### 3.2.9 *PhotometryFilter.transmissionCurve*

This data model field stores points of the curve in place in a simple table using spectrum data fields as shown above. See serialization example in Appendix C section 6.1

#### 3.2.9.1 PhotometryFilter.transmissionCurve.access

If the transmission curve is hooked as an external file, we use the *Access* class defined in the Observation CoreComponents data model [11] and inherited from the SSA specification [12]

##### *3.2.9.1.1 PhotometryFilter.transmissionCurve.access.reference*

The access reference is a URI (typically a URL) which can be used to retrieve the specific dataset described in a row of the query table response.

##### *3.2.9.1.2 PhotometryFilter.transmissionCurve.access.format*

The PhotometryFilter.transmissionCurve.access.format data model field tells the MIME type of the file pointed to and used to store the curve points. Values for this string can generally be:

```
application/fits
application/x-votable+xml
text/csv
text/xml
```

The file content will be a spectrum serialization with *PhotometryFilter.transmissionCurve.spectrum.Dataset.DataModel* set to "Spectrum1.1" for instance , and all necessary fields for the spectral and flux coordinates.

##### *3.2.9.1.3 PhotometryFilter.transmissionCurve.access.size*

Approximate estimated size of the dataset, specified in kilobytes. This would help the client estimate download times and storage requirements when generating execution plans. Only an approximate, order of magnitude value is required (a value rounded up to the nearest hundred kB would be sufficient).

#### 3.2.9.2 PhotometryFilter.transmissionCurve.transmissionPoint

The transmission curve is a mathematical function that describes the transmission fraction of a certain filter in a defined spectral range. This function can be discretized as a set of transmission points and every point will be composed by two attributes:

- One spectral coordinate (wavelength, energy or frequency) value, of type PhysicalQuantity, and utype:

***photDM:PhotometryFilter.transmissionCurve.transmissionPoint.spectralValue***

- One transmission unitless value between 0 and 1 of type double and utype:

***photDM:PhotometryFilter.transmissionCurve.transmissionPoint.transmissionValue***

#### 3.2.9.2.1 *PhotometryFilter.transmissionCurve.transmissionPoint.spectralValue.UCD: String*

This data model field contains a Unified Content Description string (UCD) [6] that specifies the nature of the spectral axis for this filter. This applies to the full spectral axis description of the filter.

Example:

```
em.wl
```

Where *em.wl* indicates that the spectral coordinate is provided in wavelength.
The Unit and UCD strings follow specific constraints defined in the IVOA standards and are implemented using type restrictions on strings.

### 3.2.10 PhotometryFilter.bandwidth: S_Bounds

A reference position along the spectral axis coverage of the referenced photometry filter.

Although this will partially reuse the

Char.SpectralAxis.Coverage.Location.Bounds

concept of the Characterization Data Model, the basic elements of this object are described within the context of a photometry filter as follows.

#### 3.2.10.1 PhotometryFilter.bandwith.UCD: String

Unified Content Description (UCD) string that specifies the nature of the bandwidth object.

#### 3.2.10.2 PhotometryFilter.bandwith.unit: IVOA.Unit

Field that specifies the units of the bandwidth object.

#### 3.2.10.3 PhotometryFilter.bandwith.extent: double

For square filters (100% between the minimum and maximum wavelength and 0% otherwise), the bandwidth could be described as $\lambda_{\max} - \lambda_{\min}$.

However, for real filters, the bandwidth is not very usable to describe the band-pass of the filter, but the effective width, that can be described as follow:

$$w = \frac{\int T(\lambda) d\lambda}{Max(T(\lambda))}$$

where $w$ is the effective width, $T(\lambda)$ is the transmission curve (see below) and $Max(T(\lambda))$ the maximum value of the transmission curve. As in previous points, please notice that, since the transmission curve will be only defined in a specific spectral range, the integrals will also be defined in this spectral range.

#### 3.2.10.4 PhotometryFilter.bandwith.start: double

Also called $\lambda_{\min}$ in the rest of the document, this is a spectral value that better describes the minimum value of the spectral range of the filter band-pass. In general, although this will not be imposed in order to allow a better description for different types of transmission curves, this quantity will be close to:

$$\lambda_{\min} = \lambda_{mean} - \frac{w}{2}$$

In practice, this could be taken as the minimum value of the filter transmission curve.

#### 3.2.10.5 PhotometryFilter.bandwith.stop: double

Also called $\lambda_{\max}$ in the rest of the document, this is a spectral value that better describes the maximum value of the spectral range of the filter band-pass. In general, although this will not be imposed in order to allow a better description for different types of transmission curves, this quantity will be close to:

$$\lambda_{\max} = \lambda_{mean} + \frac{w}{2}$$

In practice, this could be taken as the maximum value of the filter transmission curve.

## 3.3 PhotCal Class

Class to describe the use of a photometry filter by using a certain magnitude system configuration. It has associated a certain zero point object.

### 3.3.1 PhotCal.identifier: String

This field identifies, in a unique way, within a certain Photometry Filter Profile service, a zero point assigned to a filter and a certain photometric system type. Although the main requirement of the uniqueIdentifier is to be unique within a Filter Profile Service, the suggested syntax would be:

| |
|---|
| Facility/Subcategory/Band/Photometric System Type[/Suffix] |

where *Facility* is the telescope, observatory, space mission, etc that has this filter, *Subcategory* is a meaningful classification of filters within a facility (usually instrument), *Band* is the generic name used to describe the wavelength band used by this filter *Photometric System Type* makes reference to the type of system as per classification within this document and *Suffix* is optional metadata added to the unique identifier string to ensure uniqueness within a Filter Profile Service.

Please notice the suggested syntax of PhotCal unique identifier syntax corresponds with the Photometry Filter unique identifier concatenated with the photometric system type.

Example:

```
SDSS/SDSS.G/G/AB
```

### 3.3.2 PhotCal.zeroPoint: ZeroPoint

Zero point object associated to this PhotCal instance.

### 3.3.3 PhotCal.magnitudeSystem: MagnitudeSystem

Magnitude system object associated to this phot cal instance.

## *3.4 ZeroPoint Class*

This class is used to characterize a zero point flux obtained during the calibration of a certain photometry filter on a certain photometric system configuration. This object includes references to the relevant Photometric System and Photometry Filter objects.

### 3.4.1 ZeroPoint.flux.value: double

Flux of an astronomical object that produces a magnitude of reference (usually set as zero) for this particular filter and photometric system. This quantity is necessary to convert to flux a certain magnitude.

For Pogson magnitudes (see section 3.2.5) it will be used in the following way:

$$f = f_0 \, 10^{-(m-m_R)/2.5}$$

See ZeroPoint.type description for other definitions.

The flux could be expressed as $f_\lambda$ or $f_\nu$, leaving the characterization of the type of flux to the units in which this quantity is expressed.

### 3.4.2 ZeroPoint.referenceMagnitude.value: double

Most of the time, the zero point flux is defined for a magnitude=0 value. However, to give room to other cases, another reference magnitude value can be given instead of zero. The use of this reference magnitude is described in the different getMagnitudeFromFlux() and getFluxFromMagnitude() zero point extension operations.

Please notice that, by default, reference magnitude will be zero unless specified otherwise.

Reference magnitude is a dimensionless variable. It is modeled using a PhysicalQuantityDouble object type.

### 3.4.3 ZeroPoint.referenceMagnitude.error: double

Total error estimated of the reference magnitude whenever applicable.
Reference Magnitude error is a dimensionless variable.

### 3.4.4 ZeroPoint.type: enum

Usual definition of magnitudes, also called Pogson magnitudes, can be improved for faint sources by replacing the usual logarithm with an inverse hyperbolic sine function. These kinds of magnitudes are called “asinh magnitudes” or “luptitudes” [9]

| Zero point type | Value | Description |
|---|---|---|
| Pogson | 0 (default) | Usual definition |
| Asinh | 1 | Used for faint sources, replacing the usual logarithm with an inverse hyperbolic sine function. |
| LinearFlux | 2 | Linear (not logarithmic) magnitudes used in Radio, Far Infrared, X-Ray spectral |

The main difference between the three types of zero points is the conversion formulae to be used when translating magnitudes into flux and reverse.

In the ZeroPoint class we define two conversion functions; getMagnitudeFromFlux() and getFluxFromMagnitude() defined as:

- getMagnitudeFromFlux()
  - Input Parameters: Flux given in units defined in the ZeroPoint.unit data model field.
  - Output Result: Corresponding magnitude in double.

- getFluxFromMagnitude()
  - Input Parameters: Magnitude in double
  - Output Result: Corresponding flux given in units defined in the ZeroPoint.unit data model field.

## 3.5 PogsonZeroPoint Class

Extension of ZeroPoint to accommodate standard logarithm magnitudes. It has no supplementary attributes but specific conversion functions.

### 3.5.1 PogsonZeroPoint.getFluxFromMagnitude()

Operator to convert from a flux to a magnitude for Pogson magnitudes. For Pogson magnitudes, the usual definition should be used:

$$f = f_0 \, 10^{-(m-m_R)/2.5}$$

Where $f$ is the associated flux, $f_0$ is the flux of reference, $m_0$ is the magnitude of reference (by default equals to zero) and $m$ is the observed magnitude.

### 3.5.2 PogsonZeroPoint.getMagnitudeFromFlux()

Operator to convert from a flux to a magnitude for Pogson magnitudes. For Pogson magnitudes, the usual definition should be used:

$$m = m_R - 2.5\log\left(\frac{f}{f_0}\right)$$

Where $f$ is the associated flux, $f_0$ is the flux of reference, $m_R$ is the magnitude of reference (by default equals to zero) and $m$ is the observed magnitude.

## 3.6 AsinhZeroPoint Class

Extension of ZeroPoint to describe asinh magnitudes, a.k.a. luptitudes.

### 3.6.1 AsinhZeroPoint.softeningParameter: double

Parameter used to correct the calculation of magnitudes for faint sources. Usually called b. See [13] for a formal explanation.

Example:

Values used for SDSS DR5 asinh magnitudes:

| Band | Softening Parameters (*b* coefficients) |
|---|---|
| U | $1.4 \times 10^{-10}$ |
| G | $0.9 \times 10^{-10}$ |
| R | $1.2 \times 10^{-10}$ |
| I | $1.8 \times 10^{-10}$ |
| Z | $7.4 \times 10^{-10}$ |

### 3.6.2 AsinhZeroPoint.getFluxFromMagnitude()

For asinh magnitudes, the operator to be used is:

$$f = f_0\, 10^{-(m-m_R)/2.5}\left[1 - b^2\, 10^{2(m-m_R)/2.5}\right]$$

Where $f$ is the flux of the observed source, $f_0$ is the zero point flux value, $m$ is the magnitude assigned to this source, $m_0$ is the reference magnitude (default value to zero unless specified otherwise) and a new parameter appears, $b$, called the softening parameter which is referenced in this data model as the AsihnZeroPoint.softeningParameter.

### 3.6.3 AsinhZeroPoint.getMagnitudeFromFlux()

For asinh magnitudes, the operator to be used is:

$$m = m_R - \frac{-2.5}{\ln(10)}\left[\sinh^{-1}\left(\frac{f}{2bf_0}\right) + \ln(b)\right]$$

Where $m$ is the magnitude assigned to this source, $m_R$ is the reference magnitude (default value to zero unless specified otherwise), $f$ is the flux of the observed source, $f_0$ is the zero point flux value, and a new parameter appears, $b$, called the softening parameter, which is referenced in this data model as the AsihnZeroPoint.softeningParameter.

It can be seen that Pogson and Asinh magnitudes are the same if b=0 although, numerically it is recommended to use different equations to prevent infinites. See A.1

## 3.7 LinearFluxZeroPoint Class

Extension of ZeroPoint to describe simple linear flux photometry, commonly used in Radio, Far Infrared and X-ray spectral ranges. Although not being magnitudes as such, relative linear flux measurements can be included as a special and trivial case of magnitude.

### 3.7.1 LinearFluxZeroPoint.getFluxFromMagnitude()

For Linear Flux measurements, conversion used would be a linear relation instead of a logarithmic one:

$$f = f_0 \frac{m}{m_R}$$

Where $f$ is the associated flux, $f_0$ is the flux of reference, $m_R$ is the measurement of reference (default value to one, for this type of zero points, unless specified otherwise) and $m$ is the relative observed measurement.

### 3.7.2 LinearFluxZeroPoint.getMagnitudeFromFlux()

For Linear Flux measurements, linear conversion should be used to obtain the relative observed measurement:

$$m = m_R \frac{f}{f_0}$$

Where $m$ is the relative observed measurement, $m_R$ is the measurement of reference (default value to one for this type of zero points unless specified otherwise), $f$ is the associated flux and $f_0$ is the flux of reference.

## 3.8 MagnitudeSystem Class

The main difference between magnitude systems is the reference spectrum used to evaluate the magnitudes. In some occasions, the magnitude system will have a real spectrum of an existing source to calibrate all the magnitudes. In other occasions, a synthetic spectrum will be used.

### 3.8.1 MagnitudeSystem.type: String

Photometric system type used to calculate the associated zero point. Possible values are:

| MagnitudeSystem Type |
|---|
| VEGAmag |
| ABMag |
| STMag |

The list is not exhaustive. The principal difference between these photometric systems is the reference spectrum used to calculate the zero point. See section 2.2.3 for a detailed description.

### 3.8.2 MagnitudeSystem.referenceSpectrum: URI

This describes the spectrum of an astronomical object used as reference to perform photometric calibration.

This points to a Spectrum object as defined in the IVOA spectrum data model [17]. Instead of having the whole spectrum attached, we define a link to it as referenceSpectrumURI.

This is a URL, pointing to a published IVOA resource location containing the reference spectrum used.

The value of this link can be computed or derived from the spectrum data model field spec:DataID.DataSetID for instance or re-use Curation.PublisherDID which is a unique identifier within the IVOA scope.

This mechanism offers a fully general representation of a magnitude system.

Some typical types of photometric systems are:

- VEGmag: Makes use of Vega (αLyr) as the primary calibrating star. PhotometricSystem.referenceSpectrum would be the Vega SED

- ABmag: Makes use of a reference spectrum of constant flux density per unit frequency $f_\nu$:

$$f_0^{AB} = 3.631 \times 10^{-20} \, erg \; s^{-1} \; cm^{-2} \; Hz^{-1}$$

- STmag: Introduced for the HST project, it makes use of a reference spectrum of constant flux density per unit of wavelength $f_\lambda$:

$$f_0^{ST} = 3.631 \times 10^{-9} \, erg \; s^{-1} \; cm^{-2} \; \text{Å}^{-1}$$

# 4 Use Cases

## 4.1 *Conversion from magnitude to flux, using a Filter Profile Service*

The following fields are the minimal information needed in a DAL service response (SSAP or TAP) or into a serialization of the magnitude information in a catalog in order to allow the conversion from magnitudes to fluxes if a filter profile service is used:

- It MUST have one field with Utype=”**spec:Spectrum.Data.FluxAxis.value**” and UCD="phot.mag” by measurement that includes the magnitude associated to this measurement.
  Attributes to characterize the error of the measurement like spec:Spectrum.Data.FluxAxis.Accuracy.StatError, spec:Spectrum.Data.FluxAxis.Accuracy.SysError, etc could also be present in the response.
- It MUST have one field per catalog or measurement with utype=”photdm:PhotCal.identifier” that includes the identifier within the filter profile service of the filter.

The normal workflow used by an application to do the conversion would be:

- Go to the registry to obtain registration details of the Filter Profile service, using the IVOA identifier. In particular, the service URL of the service will be used to query this service using the uniqueIdentifier.
- Query the Filter Profile Service to obtain basic information of this filter. This information would be, at least:

  - photdm:PhotCal.ZeroPoint.flux.value
  - photdm:PhotCal.ZeroPoint.flux.unit.expression
  - photdm:PhotCal.ZeroPoint.type
  - photdm:PhotometryFilter.spectralLocation.value

  And optionally, any other information that could be used for a better use of the selected data, as, e.g. the Photometry Filter related information.

Please notice that all the information of the Filter Profile Service can be overwritten either in the DAL service or in the data serialization. As an example, it could be decided that the ZeroPoint.flux to be used was not the general one for this filter within the filter profile service but the night one. In this case, this corrected value would appear in the DAL response or in the data serialization so this value, and not the one on the FPS will be used for the conversions.

Flux could be then calculated as (for Pogson magnitudes, i.e. Zeropoint.type=0 and reference magnitude = 0)

$$f = f_0 \, 10^{-m/2.5}$$

Where $f_0$ is the ZeroPoint.flux.value, $m$ is the magnitude associated to the measurement and $f$ is the associated flux. The type of flux ( $f_\lambda$ or $f_\nu$) and the associated units, although they can be indirectly deduced from the field MagnitudeSystem, will be the same as those for the ZeroPoint.flux.

In case ZeroPoint.type=1 (asinh magnitudes) the value of AsinhZeroPoint.softeningParameter.value should also be used to modify the conversion formula to:

$$f = f_0 10^{-m/2.5}\left[1 - b^2 \, 10^{2m/2.5}\right]$$

# Appendix A: Conversions

## A.1 Zero point magnitude and zero point flux

The zero point flux can also be interpreted as a magnitude in the following way. Taking the previous equation and clearing the magnitude:

$$m = -2.5\log_{10}(f / f_0) = -2.5\log_{10}(f) + 2.5\log_{10}(f_0) = -2.5\log_{10}(f) + m_R$$

Where we have defined $m_R$, zero point magnitude, as the magnitude associated to the zero point flux:

$$m_R = 2.5\log_{10}(f_0)$$

e.g. for ABmag photometric systems, the magnitudes are usually defined as:

$$m_{AB,\nu} = -2.5\log_{10}(f_\nu) - 48.6$$

Which is consistent with the definition of a zero point flux of the monochromatic $f_\nu$ flux:

$$f_R^{AB} = 3.63x10^{-20} \, erg.s^{-1}cm^{-2}Hz^{-1}$$

As:

$$m_R = 2.5\log_{10}(f_0^{AB}) = 2.5\log_{10}(3.63x10^{-20}) = -48.6$$

Other systems usually define the zero point flux as a $f_\lambda$ flux, as it is usually done by, e.g., STMag systems. For these systems, the reference flux would be a monochromatic $f_\lambda$ flux:

$$f_0^{ST} = 3.631x10^{-9} \, erg \; s^{-1} \, cm^{-2} \, \text{Å}^{-1}$$

The usual definition of magnitudes for this photometric system is:

$$m_{ST,\lambda} = -2.5 \log_{10}(f_\lambda) - 21.1$$

Which corresponds to, as in the previous example, a zero point flux of

$$f_0^{ST} = 3.631x10^{-9} \, erg \; s^{-1} \, cm^{-2} \, \text{Å}^{-1}$$

as:

$$m_R = 2.5 \log_{10}(f_0^{ST}) = 2.5 \log_{10}(3.63x10^{-9}) = -21.1$$

In the present model and in order to provide a uniform treatment for all the different photometric systems, we have used the zero point flux as the quantity to characterize the photometry filter. The type of flux ( $f_\nu$ or $f_\lambda$ ) and the units of any converted to flux magnitude would coincide with the ones used to express the zero point flux, i.e., the zero point flux contains information lost in the zero point magnitude.

## *A.2 Interrelation between Pogson and Asinh magnitudes*

It can be proved that, if b=0, Pogson and Asinh magnitudes are the same:

$$\frac{f}{f_0} = 10^{-m/2.5} \left[1 - b^2 \, 10^{2m/2.5}\right]\Big|_{b=0} = 10^{-m/2.5}$$

and

$$m = \frac{-2.5}{\ln(10)}\left[\sinh^{-1}\left(\frac{f}{2bf_0}\right) + \ln(b)\right]|_{b=0} =$$

$$\frac{-2.5}{\ln(10)}\left[\ln\left(\frac{f}{2bf_0} + \sqrt{1 + \frac{f^2}{4b^2 f_0^2}}\right) + \ln(b)\right]|_{b=0} =$$

$$\frac{-2.5}{\ln(10)}\left[\ln\left(\frac{f}{2f_0} + \sqrt{b^2 + \frac{f^2}{4f_0^2}}\right)\right]|_{b=0} =$$

$$\frac{-2.5}{\ln(10)}\left[\ln\left(\frac{f}{f_0}\right)\right] =$$

$$-2.5\log\left(\frac{f}{f_0}\right)$$

Although, as can be seen in the previous calculation, the use in code of the general asinh formula for both Pogson (b=0) and asinh (b>0) magnitudes is not recommended, as it could easily produce numerical infinites during the evaluation.

# Appendix B: Data Model Summary

| General Metadata | | | | |
|---|---|---|---|---|
| **Utype** | **UCD 1+** | **Meaning** | **Default value** | **Data type** |
| Datamodel.name | meta.id | Data Model Identification | PhotCalDM-v1.0 | string |
| **Photometric System Metadata** | | | | |
| **Utype** | **UCD 1+** | **Meaning** | **Default value** | **Data type** |
| photDM:PhotometricSystem.description | meta.note | String representation Photometric System | | string |
| photDM:PhotometricSystem.detectorType | meta.code | Type of detector (e.g energy or photon counter). Possible values defined by enumeration | 0 (Energy Counter) | int |
| **Photometry Filter General Metadata** | | | | |
| **Utype** | **UCD 1+** | **Meaning** | **Default value** | **Data type** |
| photDM:PhotometryFilter.identifer | meta.ref.ivorn | Unique identifer of filter within a Filter Profile Service (FPS) | | string |
| photDM:PhotometryFilter.fpsIdentifier | meta.ref.ivorn | IVOA identifier of the Filter Profile Service | | string |
| photDM:PhotometryFilter.name | meta.id;instr.filter | Filter Name in the instrumental configuration | | string |
| photDM:PhotometryFilter.description | meta.note | Text description of the filter band | | string |

| Photometry Filter Access Metadata | | | | |
|---|---|---|---|---|
| **Utype** | **UCD 1+** | **Meaning** | **Default value** | **Data type** |
| photDM:PhotometryFilter.transmissionCurve.access.reference | meta.ref.ivorn | URI to the effective transmission curve | | URI type |
| photDM:PhotometryFilter.transmissionCurve.access.format | meta.code | File format of the pointed transmission curve | | string |
| photDM.PhotometryFilter.transmissionCurve.transmissionPoint.spectralValue.value | em.wl | Spectral value of one element of the transmission curve representation | | double |
| photDM.PhotometryFilter.transmissionCurve.transmissionPoint.transmissionValue.value | phys.transmission | Transmission value of one element of the transmission curve representation | | double |

| Photometry Filter Spectral Axis Coverage | | | | |
|---|---|---|---|---|
| **Utype** | **UCD 1+** | **Meaning** | **Default value** | **Data type** |
| photDM:PhotometryFilter.bandName | instr.bandpass | Generic name for the filter spectral band | | string |
| photDM:PhotometryFilter.spectralLocation.value | em.wl;meta.main | Reference position along the spectral axis. Spectral coordinate of the Zero Point | | double |
| photDM:PhotometryFilter.spectralLocation.unit.expression | meta.unit | Unit of the spectral axis used to characterize it | angstrom | string |
| photDM:PhotometryFilter.spectralLocation.UCD | meta.ucd | UCD for the nature of spectral axis wl, freq, energy | em.wl | string |
| photDM:PhotometryFilter.bandwidth.unit.expression | meta.unit | Unit of the spectral extent used to | angstrom | string |
| | | characterize the bandwidth object | | |
| photDM:PhotometryFilter.bandwidth.UCD | meta.ucd | UCD for the nature of spectral bandwidth wl, freq, energy | em.wl | string |
| photDM:PhotometryFilter.bandwidth.extent.value | instr.bandwidth | Spectral axis extent of the filter | | double |
| photDM:PhotometryFilter.bandwidth.start.value | em.wl;start | Minimum value of the filter spectral coverage | | double |
| photDM:PhotometryFilter.bandwidth.stop.value | em.wl;stop | Maximum value of the filter spectral coverage | | double |
| **Photometry Filter Time Axis Coverage** | | | | |
| **Utype** | **UCD 1+** | **Meaning** | **Default value** | **Data type** |
| photDM:PhotometryFilter.dateValidityFrom | time.start | Time stamp for Start of validity for this filter in ISOTime format | | string |
| photDM:PhotometryFilter.dateValidityTo | time.end | Time stamp for Stop of validity for this filter in ISOTime format | | string |
| **PhotCal Metadata** | | | | |
| **Utype** | **UCD 1+** | **Meaning** | **Default value** | **Data type** |
| photDM:PhotCal.identifier | meta.ref.ivorn | Unique identifier of the Photometry Calibration instance within a FPS | | string |
| photDM:PhotCal.zeroPoint.flux.unit.expression | meta.unit | unit for Zero point flux | Jy | string |
| photDM:PhotCal.zeroPoint.flux.UCD | meta.ucd | ucd for Zero point flux | phot.flux.density | string |
| photDM:PhotCal.zeroPoint.flux.value | phot.flux.density | flux value at Zero point associated to this filter | | double |
| photDM:PhotCal.zeroPoint.flux.error | phot.flux.density; stat.error | Error in the flux value at Zero point associated to this filter | | double |
| photDM:PhotCal zeroPoint.referenceMagnitude.value | phot.mag | Reference magnitude used for zero point | 0.0 | double |
| photDM:PhotCal.zeroPoint.referenceMagnitude.error | phot.mag;stat.error | Error in the reference magnitude used for zero point | 0.0 | double |
| photDM:PhotCal.zeroPoint.type | meta.code | Type of zero point | 0 | int |
| photDM:PhotCal.magnitudeSystem.type | meta.code | Type of magnitude system | VEGAMag | string |
| photDM:PhotCal.magnitudeSystem.ReferenceSpectrumURI | meta.ref.ivorn | Reference SED or spectrum for this magnitude system | | uri type |
| photDM:AsinhZeroPoint.softeningParameter | obs.param | Correction parameter for luptitudes | 0.0 | double |

The proposed Utypes are defined following the IVOA rules applied for other IVOA data models and derived from a simplified XML schema.

Simplification from UMI to XML schema:
ZeroPoints may belong to one of three categories: Pogson, Asinh or LinearFlux (leaving room for other future extensions). The treatment of the different categories ZeroPoints differs from the algorithmic point of view. However, the data structure only differs in the current DM by the addition of the softening parameter attached to the Asinh case.

Transmission curves are also only considered using the Access class to a remote file in the XML schema. Other serializations using array of points are directly covered by the serialization examples.

# Appendix C: Data Model Serializations

## *C.1 Filter Profile Service Serialization*

The following serialization is an example of a response of a filter profile service making use of the Photometry Filter DM through utypes:

```
<?xml version="1.0"?>
<VOTABLE
version="1.1" xsi:schemaLocation="http://www.ivoa.net/xml/VOTable/v1.1"
 xmlns:xsi="http://www.w3.org/2001/XMLSchema-instance">

  <INFO name="QUERY_STATUS" value="OK"/>

  <RESOURCE type="results">

        <PARAM  name="fpsID"
        utype="photDM:PhotometryFilter.fpsIdentifier"
        value="ivo://svo.cab/fps"     datatype="char" arraysize="*"/>

        <PARAM  name="filID"
        utype="photDM:PhotometryFilter.identifier"
        value="2MASS/2MASS.H"         datatype="char" arraysize="*"/>

        <PARAM  name="fname"
        utype="photDM:PhotometryFilter.name"
        value="2MASS.H"               datatype="char" arraysize="*"/>

        <PARAM  name="xunit"
        utype="photDM:PhotometryFilter.spectralLocation.unit.expression"
        value="Angstrom"              datatype="char" arraysize="*"/>

        <PARAM  name="xucd"
        utype="photDM:PhotometryFilter.spectralLocation.UCD"
        value="em.wl"                 datatype="char" arraysize="*"/>
```

```
        <PARAM name="xmain"
        utype="photDM:PhotometryFilter.spectralLocation.value"
        value="16620" unit="Angstrom"  datatype="float"/>

        <PARAM  name="xwid"
        utype="photDM:PhotometryFilter.bandwidth.extent.value"
        value="2509.40236716"          datatype="float"/>

        <PARAM  name="xmin"
        utype="photDM:PhotometryFilter.bandwidth.start.value"
        value="15370"                  datatype="float"/>

        <PARAM  name="xmax"
        utype="photDM:PhotometryFilter.bandwidth.stop.value"
        value="17870"                  datatype="float"/>

        <PARAM  name="detector"
        utype="photDM:PhotometricSystem.detectorType"
        value="0"                      datatype="int"/>

        <PARAM  name="photcal"
        utype="photDM:PhotCal.identifier"
        value="2MASS/2MASS.H:Vega"     datatype="char" arraysize="*"/>

        <PARAM  name="magtype"
        utype="photDM:PhotCal.magnitudeSystem.type"
        value="VEGAmag"                        datatype="float"/>

        <PARAM name="zpvalue"
        utype="photDM:PhotCal.zeroPoint.flux.value"
        value="1024" unit="Jy"         datatype="float"/>

        <PARAM name="zpunit"
        utype="photDM:PhotCal.zeroPoint.flux.unit.expression"
        value="Jy"                     datatype="char" arraysize="*"/>

        <PARAM  name="zptype"
        utype="photDM:PhotCal.zeroPoint.type"
        value="0"                      datatype="float"/>

        <PARAM name="Facility"
        ucd="instr.obsty" value="2MASS" datatype="char" arraysize="*"/>

        <PARAM name="ProfileReference"
value="http://www.ipac.caltech.edu/2mass/releases/allsky/doc/sec6_4a.html"
        datatype="char" arraysize="*"/>

        <PARAM name="CalibrationReference"
value="http://adsabs.harvard.edu/cgi-bin/nph-bib_query?bibcode=2003AJ....126.1090C"
        datatype="char" arraysize="*"/>
```

```
<!—here the points of the response curve are stored directly in an
attached table→
    <TABLE utype="photDM:PhotometryFilter.transmissionCurve">

        <FIELD name="Wavelength"
        utype="photDM.TransmissionPoint.spectralValue.value"
        ucd="em.wl" unit="Angstrom" datatype="float"/>

        <FIELD name="Transmission"
        utype="photDM.TransmissionPoint.transmission.value"
        ucd="phys.transmission" datatype="float"/>

        <DATA>
        <TABLEDATA>
          <TR>
            <TD>12890.0</TD>
            <TD>0.0000000000</TD>
          </TR>
          <TR>
            <TD>13150.0</TD>
            <TD>0.0000000000</TD>
          </TR>
          <TR>
            <TD>13410.0</TD>
            <TD>0.0000000000</TD>
          </TR>
          <TR>
            <TD>18930.0</TD>
            <TD>0.1</TD>
          </TR>
          <TR>
            <TD>19140.0</TD>
            <TD>0.2</TD>
        ...............................
          </TR>
        </TABLEDATA>
        </DATA>
    </TABLE>
  </RESOURCE>
</VOTABLE>
```

## C.2 Photometric Data in Cone Search

Catalogs could include photometric measurements in some columns. In order to allow the publication of these measurements in a. e.g., cone search service, the creation of a new capability has been proposed.

The workflow to make use of this capability will be as follows:

- A cone search (or a future TAP service) will be registered with a certain agreed capability, e.g., Photometry.

- The response of this service will contain some VOTable groups that make use of Photometry, Spectral and Characterization data model utypes (it could also make use of links to a Filter Profile Service).
- Client applications able to process this photometric information will first look for services with this capability and make use of the information attached in the VOTable groups to handle it, e.g. by the conversion from magnitude to fluxes.

As an example, the serialization of the 2MASS catalog in a cone search service, could have the following information in the VOTable header:

```
<GROUP name="Flux1" ucd="phot.mag" utype="spec:Data.FluxAxis.Value">
<DESCRIPTION>2MASS J magnitude.</DESCRIPTION>

<PARAM name="ID" ucd="meta.id;instr.filter"
utype="photDM:PhotometryFilter.identifier"
unit="" datatype="char" arraysize="*" value="2MASS/2MASS.J" />

<PARAM name="WavelengthMean" ucd="em.wl.effective"
utype="photDM:PhotometryFilter.spectralLocation.value" unit="Angstrom"
datatype="float" value="12410.5176673" />

<PARAM name="WavelengthMin" ucd="em.wl;stat.min"
utype="photDM:PhotometryFilter.bandwidth.start.value" unit="Angstrom"
datatype="float" value="10660" />

<PARAM name="WavelengthMax" ucd="em.wl;stat.max"
utype="photDM:PhotometryFilter.bandwidth.stop.value" unit="Angstrom"
datatype="float" value="14420" />

<PARAM name="PhotSystem" ucd=""
utype="photDM:PhotometricSystem.description" unit=""
datatype="char" arraysize="*" value="2MASS" />

<PARAM name="WidthEff" ucd="instr.bandwidth"
utype="photDM:PhotometryFilter.bandwidth.extent.value" unit="Angstrom"
datatype="float" value="1624.31986357" />

<PARAM name="ZeroPoint" ucd="phot.mag;arith.zp"
utype="photDM:PhotCal.zeroPoint.flux"
unit="Jy" datatype="float" value="1614.45260952" />

<PARAM name="Description" ucd="meta.note"
utype="photDM:PhotometryFilter.description" unit="" datatype="char"
arraysize="*"
value="2MASS J" />

<PARAM name="TransmissionCurve" ucd="DATA_LINK"
utype="photDM:PhotometryFilter.transmissionCurve.access.reference"
datatype="char" arraysize="*"
value="http://svo.laeff.inta.es//theory/fr/fps.php?ID=2MASS/2MASS.J" />

<FIELDref ref="phot_m1" ucd="phot.mag;em.IR.J"
utype="spec:Data.FluxAxis.Value" />
```

```
<FIELDref ref="phot_e1" ucd="stat.error;phot.mag;em.IR.J"
utype="spec:Data.FluxAxis.Accuracy.StatError" />
</GROUP>
```

Exact details on how to serialize the response are contained in Derriere's IVOA note [14].

# References

[1] **Bessell, M.S.** (2005) **Standard Photometric Systems**, Ann.Rev.Astron.Astrophys, 43:293-336

[2] **Longo G., Notes on photometric systems zero-points**, IVOA internal report (2008)

[3] **O'Connell, Magnitude and Color Systems**, ASTR 511/O'Connell Lec 14

[4] **Bruijne** (2003)**Stellar fluxes: transformations and calibrations**, ESA Tech. Note

[5] **Plante R et a**l (2007) **IVOA identifiers**

[6] **Preite Martínez, A** (2007) **The UCD1+ Controlled Vocabulary** , http://www.ivoa.net/Documents/latest/UCDlist.html

[7] **Oke J.B., Wade R.A., A** (1982) **Spectrophotometric Survey of Cataclysmic Variable Stars,** The Astrophysical Journal, V87, N4, 670-679

[8] **Straizys V.** (1996) **The method of Synthetic Photometry**, Baltic Astronomy, V5, 459 476

[9] **Girardi et al** (2002) **Theoretical isochrones in several photometric systems,** A&A 391, 195-212

[10] **Maiz, J.** (2007) **A Uniform Set of Optical/NIR Photometric Zero Points to be Used with CHORIZOS**, ASP Conference Series, V999

[11] **Louys M. et al** (2011) **Observation Data Model Core Components and its Implementation in the Table Access Protocol v1.0,** http://www.ivoa.net/Documents/ObsCore/

[12] **Tody D. et al** (2011) **Simple Spectral Access Protocol (SSAP) v1.1**, http://www.ivoa.net/Documents/SSA/20110417/PR-SSA-1.1-20110417.pdf

[13] **Lupton, Robert H.; Gunn, James E.; Szalay, Alexander S.** (1999) **A Modified Magnitude System that Produces Well-Behaved Magnitudes, Colors, and Errors Even for Low Signal-to-Noise Ratio Measurements**, The Astrophysical Journal, V118, Issue 3, pp. 1406-1410

[14] **Derriere, S** (2010) **Providing Photometric Data Measurements Description in VOTables,** http://www.ivoa.net/internal/IVOA/PhotometryDataModel/NOTE-PPDMDesc-0.1-20101202.pdf

[15] **Murdin, P** (2000) **Encyclopedia of Astronomy and Astrophysics, p 1939**

[16] **Bohlin, R. C,,Gilliland, R. L.** (2004) **Hubble Space Telescope Absolute Spectrophotometry of Vega from the Far-Ultraviolet to the Infrared,** The Astronomical Journal, Volume 127, Issue 6, pp. 3508-3515

[17] **McDowell J. et al** (2011) **IVOA Spectral Data Model v1.1,** http://www.ivoa.net/Documents/SpectrumDM/index.html